\documentclass{article}
\usepackage{spconf}

\usepackage[T1]{fontenc}
\usepackage{graphicx}
\usepackage{cite}
\usepackage{subcaption}

\usepackage{overpic}
\usepackage{amssymb}
\usepackage{amsmath}
\usepackage{amsthm}
\usepackage{array}
\usepackage{color}
\usepackage{url}

\theoremstyle{definition}

\newtheorem{example}{Example}

\newtheorem{proposition}{Proposition}

\newcommand{\vect}[1]{\mathbf{#1}}

\def\diag{\mathrm{diag}}

\def\Ttran{\mbox{\tiny $\mathrm{T}$}}
\def\CN{\mathcal{N}_{\mathbb{C}}} 

\def\Ptx{P_{\mathrm{tx}}}

\title{Demystifying the Power Scaling Law of \\ Intelligent Reflecting Surfaces and Metasurfaces\vspace{-0.1cm}}
\twoauthors
  {\vspace{-0.4cm}Emil Bj{\"o}rnson\sthanks{E. Bj{\"o}rnson is supported by ELLIIT and the Swedish Research Council.}}
	{\small Department of Electrical Engineering (ISY)\\ \small$\!\!\!$Link\"{o}ping University, 58183 Link\"{o}ping, Sweden$\!\!\!$\\ \small emil.bjornson@liu.se\vspace{-0.3cm} }%
  {\vspace{-0.4cm}Luca Sanguinetti\sthanks{L. Sanguinetti is supported by the University of Pisa under the PRA 2018-2019 Research Project CONCEPT.}}
	{$\!\!\!\!\!\!\!\!\!$\small Dipartimento di Ingegneria dell'Informazione$\!\!\!$\\ \small University of Pisa, 56122 Pisa, Italy \\\small luca.sanguinetti@unipi.it\vspace{-0.3cm} }

\begin{document}
\ninept
\maketitle

\begin{abstract}
Intelligent reflecting surfaces (IRSs) have recently attracted the attention of communication theorists as a means to control the wireless propagation channel. It has been shown that the signal-to-noise ratio (SNR) of a single-user IRS-aided transmission increases as $N^2$, with $N$ being the number of passive reflecting elements in the IRS. This has been interpreted as a major potential advantage of using IRSs, instead of conventional Massive MIMO (mMIMO) whose SNR scales \emph{only} linearly in $N$. This paper shows that this interpretation is incorrect. We first prove analytically that mMIMO always provides higher SNRs, and then show numerically that the gap is substantial; a very large number of reflecting elements is needed for an IRS to obtain  SNRs comparable to mMIMO.
\end{abstract}


\section{Introduction}

Massive MIMO (mMIMO) is the key physical layer technology in 5G \cite{Parkvall2017a}. In a nutshell, mMIMO uses a base station with many antennas (e.g., $\geq 64$) to deliver large beamforming gains and perform spatial multiplexing of many users on the same time-frequency resource \cite{Marzetta2010a,massivemimobook,Sanguinetti2019az}. In this way, the spectral efficiency can be increased by, at least, an order of magnitude compared to 4G. Since the theory and properties of mMIMO are now rather well understood \cite{massivemimobook,Sanguinetti2019az}, the research community is currently searching for what lies beyond. Five potential research directions were recently outlined in \cite{Bjornson2019d}.

One of these research directions relies on the use of an \emph{intelligent reflecting surface (IRS)} \cite{Wu2018a}, also known as \emph{reconfigurable reflectarray} \cite{Huang2005a}, \emph{software-controlled metasurface} \cite{Liaskos2018a,Renzo2019a}, and \emph{reconfigurable intelligent surface} \cite{Basar2019a}.
The key idea is to utilize a base station with a small number of antennas and support it with an IRS, deployed at another location, which takes the signal that reaches it and passively beamforms it towards the base station in the uplink and towards the user in the downlink. The IRS consists of an array of many diffusely reflecting elements \cite{Liang2015a}, which can be viewed as passive antennas since the sub-wavelength form factor is similar. Each element assigns a phase-shift to its reflected signal, so the joint effect of all elements is a reflected beam in a desired direction \cite{PhysRevX.8.011036}. The physics are the same as for beamforming with a phased array, except that the array then generates the signal locally.
The passive operation is conceptually appealing since one can (to some extent) control the wireless propagation channel, in addition to controlling the transmitter and receiver as in conventional systems. However, the performance benefits compared to conventional relays are disputable \cite{Bjornson2019e}.

Several recent works have emphasized that an IRS achieves a better power scaling law than conventional mMIMO \cite{Wu2018a,Wu2019a,Basar2019a}. The signal-to-noise ratio (SNR) in mMIMO is proportional to the number of antennas $N$  \cite{Ngo2013a,Hoydis2013a}. This implies that the transmit power needed to achieve a target SNR value during data transmission reduces as $1/N$, which is the so-called power scaling law.\footnote{If one also reduces the transmit power in the channel acquisition phase, the power scaling law changes; we refer to \cite{Ngo2013a,Hoydis2013a,massivemimobook} for details.} In contrast, \cite{Wu2018a}  showed that the SNR is proportional to $N^2$ when using an IRS, where $N$ denotes the number of reflecting elements. This seemingly mysterious scaling difference has been interpreted as a major advantage of IRS-aided communications. The natural question is:
\vspace{-0.1cm}
\begin{center}
\emph{How can an IRS with $N$ passive reflecting elements be more power-efficient than mMIMO with $N$ active antennas?}
\end{center}
\vspace{-0.1cm}
This paper takes a close look at this question, with the purpose of demystifying the interpretation mentioned above. We prove that, despite the scaling difference, an IRS can never outperform mMIMO; in fact, a substantial SNR gap is typically observed.

\section{Preliminaries}
\label{sec:preliminaries}

\begin{figure}[t!]\vspace{-0.2cm} 
        \centering
        \begin{subfigure}[t]{\columnwidth} \centering 
	\begin{overpic}[width=0.68\columnwidth,tics=10]{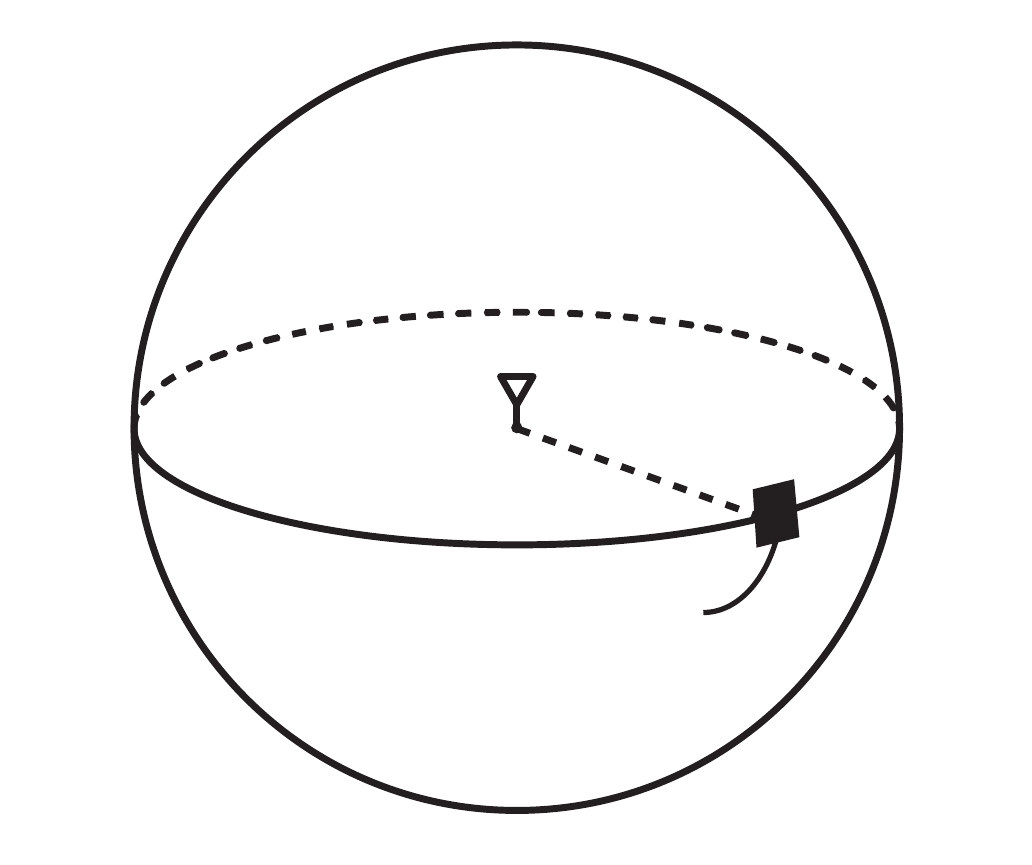}
  \put (21,43) {\small Transmitter}
 \put (62,40) {\small $d$}
  \put (28,23) {\small Receive antenna}
     \put (28,17) {\small with area $A$}
\end{overpic} \vspace{-2mm}
                \caption{One receive antenna.} 
                \label{fig:sphere}\vspace{-0.4cm}
        \end{subfigure} 
        \begin{subfigure}[t]{\columnwidth} \centering  \vspace{+2mm}
	\begin{overpic}[width=0.68\columnwidth,tics=10]{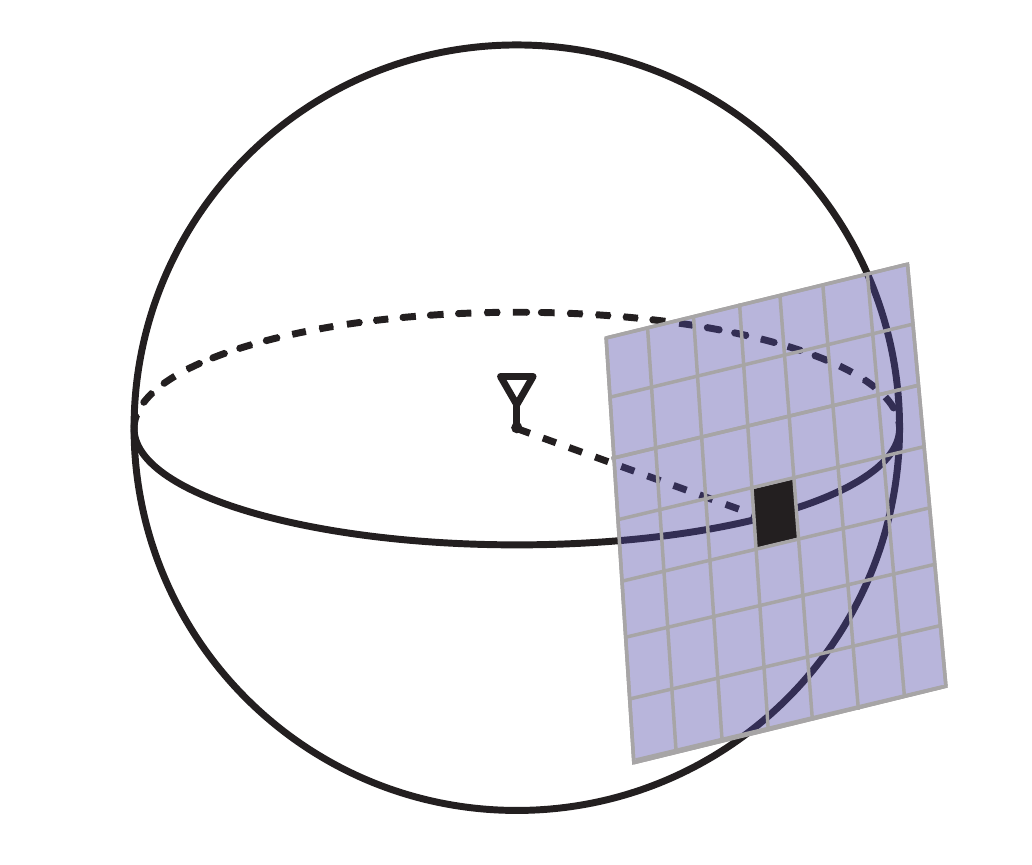}
  \put (21,43) {\small Transmitter}
 \put (62,40) {\small $d$}
  \put (21,22) {\small Planar array with}
     \put (27,16) {\small $N$ antennas}
\end{overpic}  \vspace{-2mm}
                \caption{Planar array with $\sqrt{N} \times \sqrt{N}$ receive antennas.\vspace{-0.1cm}} 
                \label{fig:sphere2}
        \end{subfigure} 
        \caption{Examples of basic propagation scenarios.}
        \label{fig:sphere12_examples} \vspace{-0.2cm} 
\end{figure}
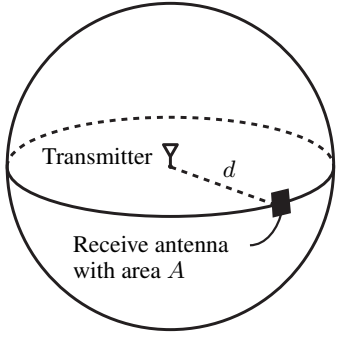
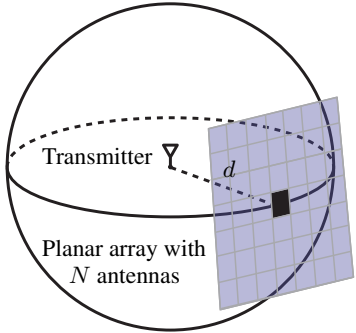

To explain the power scaling laws related to an IRS, we first need to review the basics of wireless propagation. We consider the free-space propagation scenario depicted in Fig.~\ref{fig:sphere}. An isotropic antenna transmits a signal with power $\Ptx $ and a receive antenna is located at distance $d$ from the transmitter. Suppose the receive antenna is lossless and has an (effective) area $A$ perpendicularly to the direction of propagation; that is, the receive antenna occupies an area $A$ on the surface of a sphere with radius $d$. The received power is
\begin{equation} \label{eq:received_power_SISO}
P_{\mathrm{rx}} =  \frac{A}{4\pi d^2} \Ptx = \beta \Ptx
\end{equation}
where $\beta=\frac{A}{4\pi d^2}$ denotes the free-space channel gain (pathloss) and $4\pi d^2$ is the total surface area of the sphere. The law of energy conservation states that the received power can never be higher than the transmitted power. This implies that $\beta \in [0,1]$ in \eqref{eq:received_power_SISO}. 

\begin{example}
To quantify the practical values of $\beta$, suppose the system operates at a carrier frequency of $f=3$ GHz over distances $d\in [2.5, 25]$\,m. Consider a lossless isotropic antenna with aperture/area $A= {\lambda^2}/{(4\pi)}$ where $\lambda = c/f$\,m is the wavelength and $c$ is the speed of light. In this case, $\beta$ ranges from $-40$\,dB to $-60$\,dB. \vspace{-0.2cm} \end{example}

\subsection{Spherical Antenna Arrays}

To receive more power than in \eqref{eq:received_power_SISO}, we can deploy additional antennas on the sphere in Fig.~\ref{fig:sphere}. For example, if $N$ non-overlapping antennas (of the same kind) are placed on the sphere, the collected power is $N$ times the value in \eqref{eq:received_power_SISO}:
\begin{equation} \label{eq:received_power_SISO2}
P_{\mathrm{rx}}^{\textrm{sphere-}N} = N P_{\mathrm{rx}} = N \beta \Ptx.
\end{equation}
We refer to $N \beta$ as the total channel gain and note that it grows linearly with $N$.
This fact is typically used in multiple antenna technologies to demonstrate that the received signal power grows proportionally to the number of antennas. However, due to the law of energy conservation, \eqref{eq:received_power_SISO2} is valid only if
\begin{equation}
N \beta \le 1
\end{equation}
which means that no more than $N = \frac{1}{\beta}=\frac{4\pi d^2}{A}$ antennas can be deployed; in that case, the sphere is fully covered and all the transmitted power $\Ptx$ is received. For the setup in Example~1, we need from $10^4$ to $10^6$ antennas to cover the entire sphere depending on $d$; even more antennas are needed if $d \ge 25$\,m, which is typically the case for sub-6\,GHz communication systems. If the operating frequency is of the order of $f=30$ GHz (e.g., for operation in the millimeter-wave bands where IRSs may have promising applications), then $\beta$ in Example~1 ranges from $-60$\,dB to $-80$\,dB such that a minimum number of $10^6$ antennas is required to capture all the transmitted power. These values explain why the scaling behavior in \eqref{eq:received_power_SISO2} is conventionally adopted when analyzing practical mMIMO systems, even in extreme cases with thousands of antennas. However, it must be clear that \eqref{eq:received_power_SISO2} is not applicable in the asymptotic regime where $N\to \infty$ simply because of the law of energy conservation.

\subsection{Planar Antenna Arrays}

Large antenna arrays normally have a planar rather than spherical form factor. In this case, the analysis becomes more complicated. We now consider a planar array with $N$ antennas, each having an area $A$; see Fig.~\ref{fig:sphere2}. More precisely, each antenna has size $\sqrt{A} \times \sqrt{A}$ and the array is square-shaped with $\sqrt{N} \times \sqrt{N}$ equally spaced antennas.
If the center of the array is at distance $d$ from the transmit antenna and is the closest point to the transmitter, as shown in Fig.~\ref{fig:sphere2},  then the received signal power is \cite[Eq.~(19)]{Hu2018a}
\begin{equation} \label{eq:received_power_planar}
P_{\mathrm{rx}}^{\textrm{planar-}N}  = \alpha \Ptx
\end{equation}
with 
\begin{equation}
\alpha = \frac{1}{\pi} \tan^{-1} \! \left( \frac{N A}{4d \sqrt{N A + d^2} } \right).
\end{equation}
If the array is in the far-field of the transmitter, i.e., $d \gg \sqrt{N A}$, then we have that $N A + d^2 \approx d^2$ and the argument of $ \tan^{-1}$ is close to zero. By using the first-order Taylor approximation $\tan^{-1}(x) \approx x$, we thus obtain
\begin{equation} \label{eq:received_power_planar-approx}
P_{\mathrm{rx}}^{\textrm{planar-}N}  \approx \Ptx \frac{1}{\pi} \frac{N A}{4d \sqrt{d^2} }  = N \beta \Ptx
\end{equation}
which is equal to $P_{\mathrm{rx}}^{\textrm{sphere-}N}$ in  \eqref{eq:received_power_SISO2}. Hence, for relatively small planar arrays, the received  power is proportional to $N$.

If $N$ grows large, the far-field approximation cannot be applied anymore and we instead notice that 
\begin{equation}
 \tan^{-1} \! \left( \frac{N A}{4d \sqrt{N A + d^2} } \right) \to \frac{\pi}{2} \quad \textrm{as} \,\,N \to \infty.
 \end{equation}
 Hence, the received power in \eqref{eq:received_power_planar} satisfies 
 \begin{equation}
 P_{\mathrm{rx}}^{\textrm{planar-}N} \to \frac{\Ptx}{2} \quad \textrm{as} \,\,N\to \infty. 
  \end{equation}
 The reason is that every time we increase the array size, the new receive antennas are deployed further away from the transmitter and are also less directed towards the direction of propagation, leading to a smaller effective area. As a consequence, a gradually lower power is received by the new antennas. The limit $\Ptx /2$ represents the fact that the transmitter radiates half of its power in the half-plane where the array is located. This is in agreement with the law of energy conservation, which requires $P_{\mathrm{rx}}^{\textrm{planar-}N} \leq \Ptx$ for all $N$.

\begin{figure}[t!]\vspace{-0.2cm} 
	\centering 
	\begin{overpic}[width=1.1\columnwidth,tics=10]{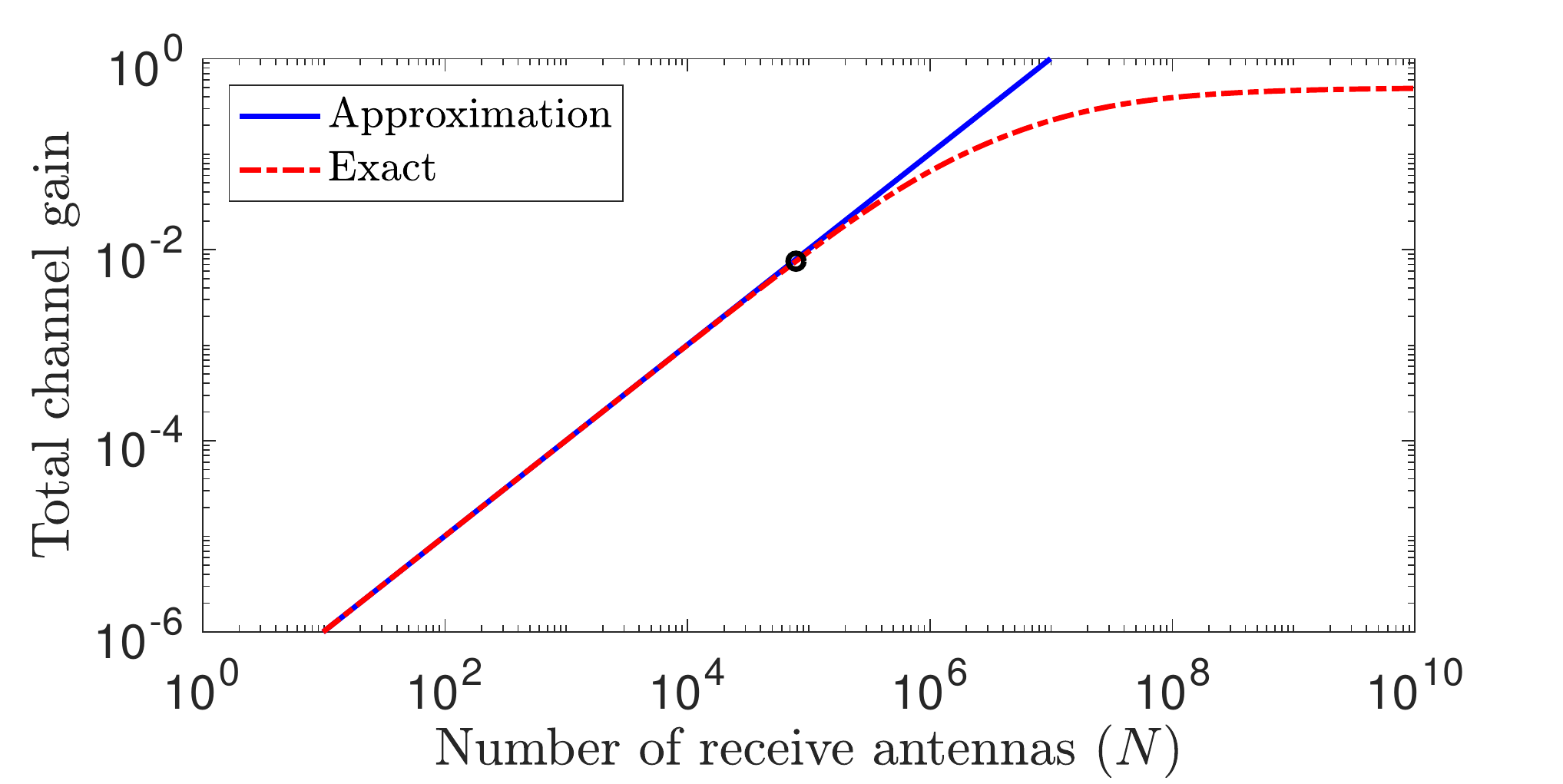}
	\put(47,25){\small Point given by the}
	\put(47,21){\small rule-of-thumb}
	\put(57,28){\vector(-1, 1){5}}
\end{overpic} 
	\caption{The total channel gain $\rho = P_{\mathrm{rx}}^{\textrm{planar-}N} / \Ptx$ with a planar array with $\sqrt{N} \times \sqrt{N}$ equally spaced antennas. The setup defined in Example~1 is considered.}
	\label{figure_propagationloss}  \vspace{-4mm}
\end{figure}

Can we utilize the approximation in \eqref{eq:received_power_planar-approx} in practice? Does the received power grow linearly with $N$ for practical array sizes?
To give a qualitative answer, Fig.~\ref{figure_propagationloss} shows the total channel gain $\rho = P_{\mathrm{rx}}^{\textrm{planar-}N} / \Ptx \in [0,1]$ as a function of $N$, using either the exact expression in \eqref{eq:received_power_planar} or the approximation in \eqref{eq:received_power_planar-approx}. We consider the setup defined in Example~1.  Fig.~\ref{figure_propagationloss} shows that nearly $10^5$ antennas are needed before the approximation error is noticeable (above 5\%), and $10^8$ antennas are needed to approach the upper limit $1/2$.
As a rule-of-thumb, the approximation is accurate for all $N$ satisfying $N A /10 < d^2$; the value of $N$ that gives equality in this rule-of-thumb is indicated by a circle in Fig.~\ref{figure_propagationloss}. As the distance $d$ increases, the number of antennas that satisfies the rule-of-thumb grows quadratically.
In conclusion, we can utilize the approximation in \eqref{eq:received_power_planar-approx} in the remainder of this paper and cover most cases of practical interest.  

\section{Power Scaling Laws}

\begin{figure}[t!]\vspace{-0.2cm} 
        \centering 
        \begin{subfigure}[b]{\columnwidth} \centering 
	\begin{overpic}[width=.95\columnwidth,tics=10]{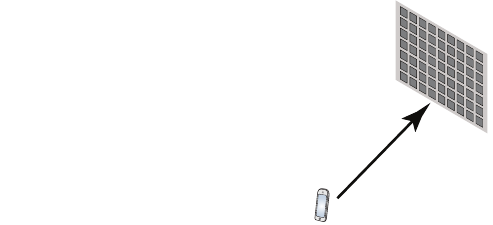}
	\put(68,0.5){\small Transmitter}
	\put(50.5,38){\small mMIMO receiver}
	\put(50.5,33){\small with $N$ antennas}
	\put(83,13.5){\small $\vect{h} \propto \sqrt{\beta_\vect{h}}$}
\end{overpic} 
                \caption{Conventional mMIMO transmission.} \vspace{3mm}
                \label{fig:mMIMOexample}
        \end{subfigure} 
        \begin{subfigure}[b]{\columnwidth} \centering  \vspace{+2mm}
	\begin{overpic}[width=.95\columnwidth,tics=10]{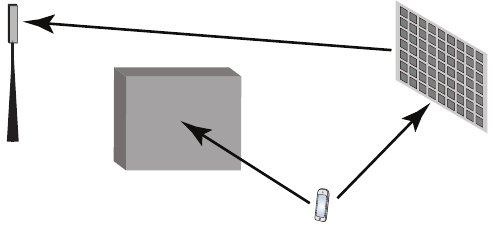}
	\put(-2,13){\small Receiver}
	\put(68,0.5){\small Transmitter}
	\put(24.5,7.5){\small Blocking object}
	\put(63,30){\small IRS with }
	\put(63,25){\small $N$ elements}
	\put(39,41){\small $\vect{g}\propto \sqrt{\beta_\vect{g}}$}
	\put(83,13.5){\small $\vect{h}\propto \sqrt{\beta_\vect{h}}$}
\end{overpic}  
                \caption{IRS-aided transmission.\vspace{-0.1cm}}  
                \label{fig:IRSexample} 
        \end{subfigure} 
        \caption{Illustration of the setups compared in this paper.\vspace{-0.3cm}} 
        \label{fig:examples} 
\end{figure}
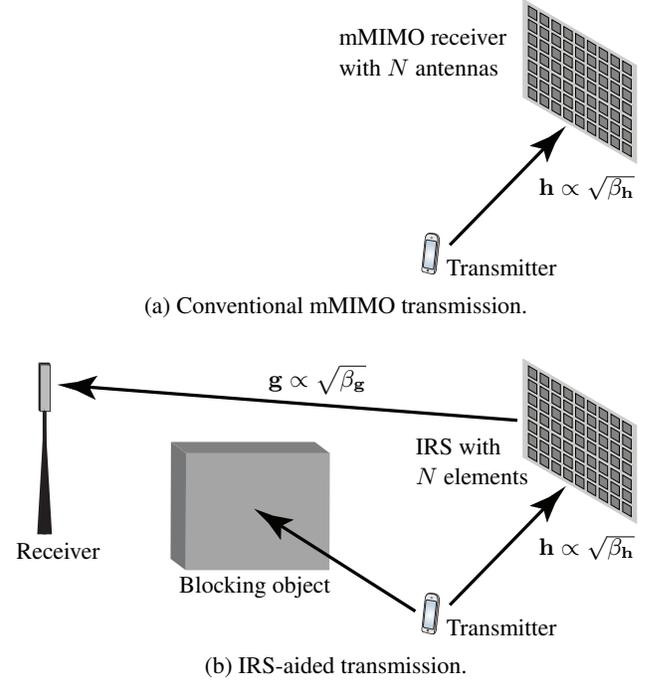

In this section, we will compare the SNR expressions and power scaling laws achieved by mMIMO and an IRS-aided transmission when the arrays are deployed at the same location. The considered mMIMO setup is illustrated in Fig.~\ref{fig:mMIMOexample}, where a single-antenna transmitter communicates with a planar array of $N$ antennas. We consider a line-of-sight scenario where the flat-fading channel is represented by the vector $\vect{h} \in \mathbb{C}^N$. In the corresponding IRS setup in Fig.~\ref{fig:IRSexample}, the mMIMO array is replaced by an IRS with $N$ passive reflecting elements. The single-antenna receiver is physically separated from the IRS and the line-of-sight channel is represented by the vector $\vect{g} \in \mathbb{C}^N$. The two channels are modeled as follows:
\begin{equation} \label{eq:channel-model}
\vect{h} = \sqrt{\beta_{\vect{h}} } \begin{bmatrix}
e^{j\phi_1}\vspace{-0.2cm} \\ \vdots \\ e^{j\phi_N} \end{bmatrix}, \quad
\vect{g} = \sqrt{\beta_{\vect{g}} } \begin{bmatrix} e^{j\psi_1} \vspace{-0.2cm}\\ \vdots \\ e^{j\psi_N} \end{bmatrix}
\end{equation}
where $\phi_1,\ldots, \phi_N,\psi_1,\ldots, \psi_N$ are arbitrary phase shifts and $\beta_{\vect{h}},\beta_{\vect{g}} \in [0,1]$ are the corresponding channel gains. In free-space propagation, the channel gains are computed as 
\begin{equation} \label{eq:pathloss-model}
\beta_{\vect{h}} =\frac{A}{4\pi d_{\vect{h}}^2}, \quad
\beta_{\vect{g}} =\frac{A}{4\pi d_{\vect{g}}^2}
\end{equation}
where $d_{\vect{h}}$ and $ d_{\vect{g}}$ are the distances, while $A$ is the area of an antenna or element. We assume that perfect channel knowledge is available in both setups, which is reasonable since the channels are deterministic. Despite simple, the two setups in Fig.~\ref{fig:examples} are sufficient to answer the question posed in the introduction and demonstrate why conventional mMIMO is always a better setup from a performance perspective. More precisely, it allows us to quantify the SNR loss by deploying at a certain location an IRS with passive elements, instead of an mMIMO array with active elements.

\subsection{Massive MIMO}
We first consider the uplink mMIMO setup where $\vect{v} \in \mathbb{C}^N$ denotes the receive combining vector.
The received signal  is $\vect{h} \sqrt{\Ptx} s +  \vect{n}  \in \mathbb{C}^N$, which after receive combining becomes
\begin{equation}
y_{\textrm{mMIMO}} = \vect{v}^{\Ttran} \vect{h} \sqrt{\Ptx} s +  \vect{v}^{\Ttran} \vect{n}
\end{equation}
where $\Ptx$ is the transmit power, $s$ is the unit-norm information signal, and $\vect{n} \sim \CN(\vect{0},\sigma^2 \vect{I}_N)$ is the receiver noise. Under the assumption of perfect channel knowledge, the information rate is $\log_2(1+\mathrm{SNR}_{\textrm{mMIMO}} )$ \cite{massivemimobook}, where
\begin{equation}
\mathrm{SNR}_{\textrm{mMIMO}} = \frac{|\vect{v}^{\Ttran}\vect{h}|^2 \Ptx}{ \| \vect{v}\|^2 \sigma^2}
\end{equation}
is the SNR. It is well-known that this SNR is maximized by \emph{maximum ratio combining} for which $\vect{v} = \vect{h}^*/\|\vect{h} \|$ \cite{massivemimobook}. This leads to
\begin{equation} \label{eq:SNR-mMIMO}
\mathrm{SNR}_{\textrm{mMIMO}} = \frac{ \|\vect{h} \|^2 \Ptx}{\sigma^2} =  \frac{ N \beta_{\vect{h}} \Ptx}{\sigma^2}
\end{equation}
where the last step utilizes the channel model in \eqref{eq:channel-model}. We notice that the SNR is proportional to $N$. Hence, when $N$ increases, the system can either benefit from a linearly increasing SNR or reduce $\Ptx$ as $1/N$ to keep the SNR constant. This is the conventional power scaling law for mMIMO \cite{Ngo2013a,Hoydis2013a}. 
Although it is popular to study the asymptotic regime where $N \to \infty$ in the mMIMO literature, we recall from Section~\ref{sec:preliminaries} that $N \beta_{\vect{h}}$ can never be larger than 1. As illustrated in Fig.~\ref{figure_propagationloss}, this is not an issue for practical propagation distances and values of $N \le 10^5$, for which one can safely say that \eqref{eq:SNR-mMIMO} is valid. But if one truly wants to let $N \to \infty$, the more accurate model in \eqref{eq:received_power_planar} must be used to develop an asymptotically accurate power scaling.

\subsection{Intelligent Reflecting Surface}

The IRS is intelligent in the sense that each of the $N$ reflecting elements can control the phase of its diffusely reflected signal. As proved in \cite{Ozdogan2019a}, the received signal $y_{\textrm{IRS}} \in \mathbb{C}$ can then be modeled as 
\begin{equation}
y_{\textrm{IRS}}  = \vect{g}^{\Ttran} \vect{\Theta} \vect{h} \sqrt{\Ptx} s + n
\end{equation}
where $\Ptx$ and $s$ are the same as in the mMIMO setup and $n \sim \CN(0,\sigma^2 )$ is the noise at the receiver. The reflection properties are determined by the diagonal matrix
\begin{equation}
\vect{\Theta} = \mu \cdot \diag\left( e^{-j \theta_1}, \ldots, e^{-j \theta_N} \right)
\end{equation}
where $\mu \in (0,1]$ is a fixed amplitude reflection coefficient and $\theta_1,\ldots,\theta_N$ are the phase-shift variables that can be optimized by the IRS based on $\vect{g}$ and $\vect{h}$. Since perfect channel knowledge is available, the resulting information rate is $\log_2(1+\mathrm{SNR}_{\textrm{IRS}} )$ \cite{Bjornson2019e}, where
\begin{equation}
\mathrm{SNR}_{\textrm{IRS}} = \frac{ \mu^2 | \vect{g}^{\Ttran} \vect{\Theta} \vect{h} |^2 \Ptx}{\sigma^2}
\end{equation}
is the SNR at the receiver. The SNR is maximized by $\theta_n = \phi_n + \psi_n$ for $n=1,\ldots,N$ \cite{Wu2018a,Bjornson2019e}, which makes all the terms in the product $\vect{g}^{\Ttran} \vect{\Theta} \vect{h}$ positive and thereby add constructively. Using \eqref{eq:channel-model}, we obtain
\begin{equation} \label{eq:SNR:IRS}
\mathrm{SNR}_{\textrm{IRS}} = \frac{ \mu^2\left| \sum\limits_{n=1}^{N} \sqrt{\beta_{\vect{g}} \beta_{\vect{h}} } \right|^2 \Ptx}{\sigma^2} =  \frac{ \mu^2 N^2 \beta_{\vect{g}} \beta_{\vect{h}} \Ptx}{\sigma^2}.
\end{equation}
Interestingly, this SNR grows quadratically with $N$, which is a faster growth rate than in the case of mMIMO in \eqref{eq:SNR-mMIMO}. This fact has been recognized in several recent works \cite{Wu2018a,Wu2019a,Basar2019a}, which have described the $\mathcal{O}(N^2)$-scaling as a reason for IRS being ``\emph{more efficient than conventional mMIMO}''  \cite{Wu2018a}.
However, when comparing the complete SNR expressions, the claimed advantage actually turns into a disadvantage. The SNR in \eqref{eq:SNR:IRS} can be factorized into two terms:
\begin{equation} \label{eq:SNR_IRS}
\mathrm{SNR}_{\textrm{IRS}}  = \hspace{-1.2cm} \!\!\!\!\!\!\overbrace{ \vphantom{\frac{  N\beta_{\vect{h}} \Ptx}{\sigma^2}} \mu^2 N  \beta_{\vect{g}}}^{\leq 1, \textrm{ Fraction of reflected power reaching receiver}}  \!\!\!\!\hspace{-1.4cm}\times \,\, \hspace{-0.2cm}\underbrace{\frac{  N\beta_{\vect{h}} \Ptx}{\sigma^2}}_{=\,\mathrm{SNR}_{\textrm{mMIMO}} }.
\end{equation}
The first term represents the fraction of the power received at the IRS that also reaches the receiver. It can be further divided into the total channel gain $N  \beta_{\vect{g}}$ between the receiver and IRS, and $\mu^2$, which is the fraction that is reflected by the IRS. None of these two terms can be larger than one; in particular, Section~\ref{sec:preliminaries} described that $N  \beta_{\vect{g}}$ is an accurate total channel gain only for a sufficiently small IRS and that its value is fundamentally upper bounded by 1, due to the law of energy conservation. The second term in \eqref{eq:SNR_IRS} is the SNR in \eqref{eq:SNR-mMIMO} that was achieved by mMIMO. Therefore, we have established the following main result.

\begin{proposition}\textit{Under optimal operation, for any given $N$, it holds that}
\begin{equation}
\mathrm{SNR}_{\textrm{mMIMO}} \geq \mathrm{SNR}_{\textrm{IRS}}.
\end{equation}
\end{proposition}
Proposition 1 implies that an IRS-aided transmission can never achieve a higher rate than the corresponding mMIMO system. One way to interpret this result is that the IRS acts as an uplink mMIMO receiver that uses suboptimal receive combining $\vect{v} = \vect{\Theta} \vect{g}$, which is not the optimal maximum ratio combining, and incurs an additional SNR loss of a factor 
\begin{equation}
\| \vect{v} \|^2 = \|  \vect{\Theta} \vect{g} \|^2 \!=\! \mu^2 \| \vect{g} \|^2 \!=\! \mu^2\beta_{\vect{g}} N \leq 1.
\end{equation}

\subsection{Numerical Comparison}
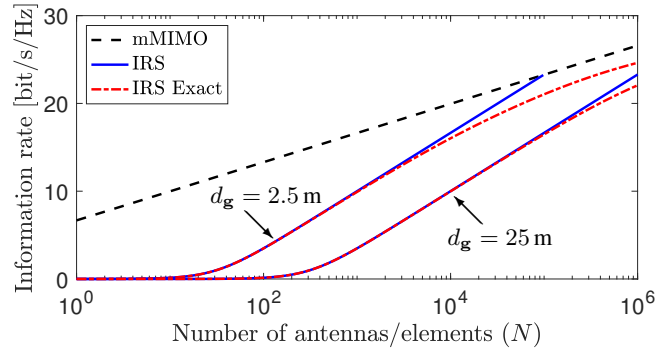
\begin{figure}[t!]\vspace{-0.2cm} 
	\centering 
	\begin{overpic}[trim=20 0 0 0,width=1.08\columnwidth,tics=10]{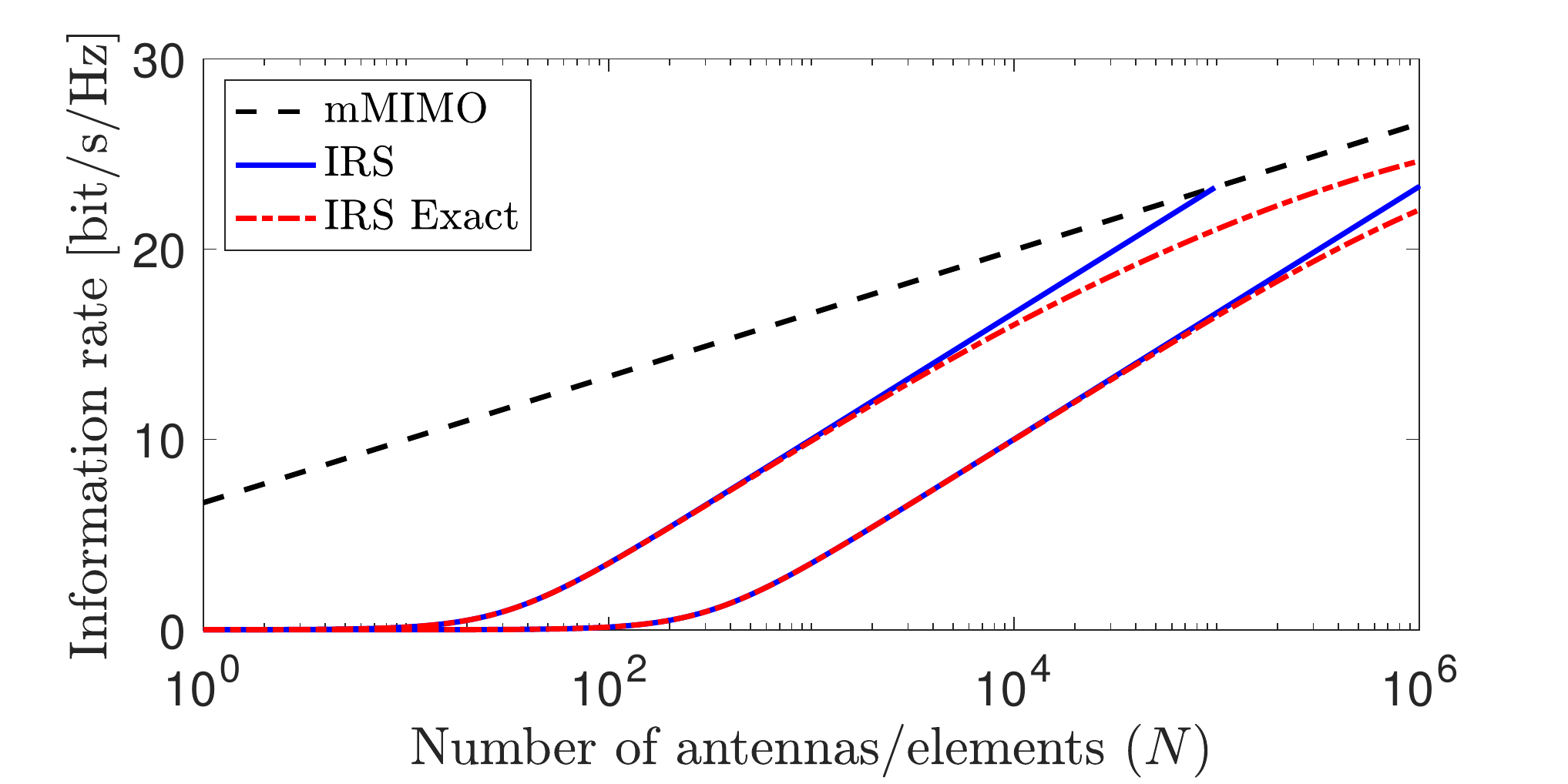}
		\put(63,15){\small $d_{\vect{g}}=25$\,m}
	\put(68,18){\vector(-1, 1){4}}
			\put(29,21){\small $d_{\vect{g}}=2.5$\,m}
	\put(34,20){\vector(1, -1){4}}
\end{overpic} 
	\caption{Information rates achieved in the mMIMO and IRS-aided setups in Fig.~\ref{fig:examples} when $d_{\vect{h}}=25$\,m, and $d_{\vect{g}}=2.5$\,m or $25$\,m.}
	\label{figure_rate_comparison}  \vspace{-0.4cm}
\end{figure}

To quantify the information rates achieved by the mMIMO and IRS-aided setups in Fig.~\ref{fig:examples}, we assume that the operating frequency is $3$\,GHz, and that each antenna/element has an area $A= {\lambda^2}/{(4\pi)}$. The distance $d_{\vect{h}}$ is fixed to $25$\,m, the transmit power is $\Ptx=10$\,mW, whereas the noise power is $\sigma^2 = 10^{-8}$\,W. Hence, we have that ${ \beta_{\vect{h}} \Ptx}/{\sigma^2} = 20$\,dB. We also assume that $\mu=1$ (i.e., no reflection loss). Fig.~\ref{figure_rate_comparison} shows the information rates as a function of the number of antennas/elements. Two different scenarios are considered. The first assumes $d_{\vect{g}}= 25$\,m such that the receiver is relatively far from the IRS; this is what it is typically conceived for IRS-aided communication \cite{Wu2019a,Liaskos2018a,Basar2019a}. The results show that the information rate achieved by the IRS has a faster growth rate than that achieved by mMIMO, since the SNRs grow as $N^2$ and $N$, respectively. Nevertheless, mMIMO provides much higher rates for any value of $N$. To achieve the same rate as mMIMO with $64$ antennas (as deployed in commercial 5G networks \cite{Sanguinetti2019az,Bjornson2019d}), an IRS-aided system needs more than $10^{4}$ reflecting elements. As observed in \cite{Bjornson2019e}, the reason is that the IRS acts as a relay that forwards the signal from the transmitter to the receiver, but without amplifying it. This results in a very low fraction of received power when the receiver is far from the IRS. 

Reflectarrays are traditionally deployed in the vicinity of a fixed receiver that is equipped with a high-gain antenna \cite{Huang2005a}, which limits the loss between the surface and  receiver. To further investigate this aspect when deploying the IRS, we also consider the case in which $d_{\vect{g}}=2.5$\,m. Recall that the channel model for $\bf g$ in \eqref{eq:channel-model} is only accurate for $N < 10d_{\vect{g}}^2/A$, which roughly limits $N$ to be below $7 \times 10^4$ when $d_{\vect{g}}=2.5$\,m. Hence, the more accurate model in \eqref{eq:received_power_planar} must  be also considered for the channel from the IRS to the receiver, which yields (by following the same arguments that led to \eqref{eq:SNR:IRS})
\begin{equation} \label{eq:SNR:IRS_exact}
\mathrm{SNR}_{\textrm{IRS}}^{\textrm{Exact}} =  \frac{ \mu^2 N \alpha_{\vect{g}} \beta_{\vect{h}} \Ptx}{\sigma^2}
\end{equation}
with
\begin{equation}
\alpha_{\bf g} = \frac{1}{\pi} \tan^{-1} \! \left( \frac{N A}{4d_{\bf g} \sqrt{N A + d_{\bf g}^2} } \right).
\end{equation}
Fig.~\ref{figure_rate_comparison} shows that the performance gap to mMIMO reduces when the receiver is closer to the IRS, but as much as $3\times10^{3}$ reflecting elements are needed in the IRS to obtain the same performance as mMIMO with $64$ antennas. The approximation error becomes noticeable when $N > 10^4$ and the error grows as we approach $10^{6}$ antennas/elements. Note that we stopped plotting the approximate IRS curve where it erroneously reached the mMIMO curve at $N\!=\!1/  \beta_{\vect{g}}$.

\section{Conclusion}
This paper compared conventional mMIMO with IRS-aided communications, when the arrays are deployed at the same location. We proved analytically that the latter can never provide higher information rate or SNR---even if the SNR scales proportionally to $N^2$ when using an IRS and only proportionally to $N$ in mMIMO \cite{Wu2018a}. The reason is that one $N$-term accounts for the fraction of power that is lost in the IRS's reflection.
Numerical results showed that thousands of reflecting elements are needed in an IRS-aided system to achieve rates comparable to mMIMO. As previously pointed out in \cite{Bjornson2019e}, this is due to the fact that the signal is forwarded by the IRS from the transmitter to receiver without any amplification. This results in a very low fraction of received power, even when the receiver is in the vicinity of the IRS, as demonstrated in Fig.~\ref{figure_rate_comparison}.

One way to alleviate this power loss is to place the IRS's receiver right behind the surface and ``reflect'' the signals to it through the surface. This is done in so-called \emph{holographic beamforming} \cite{Black2017,Bjornson2019d} and such an implementation can theoretically provide the same performance as mMIMO, while potentially being more power efficient.

\bibliographystyle{IEEEtran}

\bibliography{IEEEabrv,refs}

\end{document}